\documentstyle[sprocl,psfig]{article}

\input epsf.sty

\newcount\fcount\fcount=0

\def\ref#1{\global\advance\fcount by 1 
\global\xdef#1{\relax\the\fcount}}

\def\today{\ifcase\month\orJanuary\or February\or 
March\or April\or May\or June\orJuly\or August\or 
September\or October\or November\or 
December\fi\space\number\day, \number\year}

\def\pp{\parshape 2 0truecm 15truecm .5truecm 
14.5truecm}
\def\ref #1;#2;#3;#4{\par\pp #1, {\it #2}, {\bf #3}, 
#4}
\def\book #1;#2;#3{\par\pp #1, {\it #2}, #3}
\def\rep #1;#2;#3{\par\pp #1, #2, #3}

\def\simlt{\lower.5ex\hbox{$\; \buildrel < \over \sim \;$}}
\def\simgt{\lower.5ex\hbox{$\; \buildrel > \over \sim \;$}}
\def\simpropto{\lower.2ex\hbox{$\; \buildrel \propto \over \sim \;$}}
\def\frac #1#2{{#1\over #2}}
\def\deg{\ifmmode^\circ\;\else$^\circ\;$\fi}

\def\arcmin{\ifmmode^\prime\;\else$^\prime\;$\fi}
\def\arcsec{\ifmmode^{\prime\prime}\;\else$^{\prime\prime}\;$\fi}

\begin{document}
\title {From the Cosmological Microwave Background 
to Large-Scale Structure}
\author{Joseph Silk and Eric Gawiser}
\address{ Departments of Astronomy and Physics, and Center for Particle
Astrophysics, \\
University of California, Berkeley, CA 94720}
\maketitle
\abstracts{
The shape of the primordial
fluctuation spectrum is  probed
by cosmic microwave background fluctuations
which  measure density fluctuations at $z \sim
1000$ on scales of hundreds of Mpc and from galaxy redshift surveys,
which measure   structure at low redshift out to
several hundred Mpc. The currently acceptable  library of cosmological models is
inadequate to account for the current data,
and more exotic models must be sought.
New data sets such as SDSS and 2DF are 
urgently needed to verify whether the shape discrepancies in $P(k)$ will
persist.
}
\section{Introduction}

Our understanding of primordial fluctuations in the early
universe was revolutionized first with inflation and then
by the actual detection of temperature fluctuations in the
cosmic microwave background
(CMB).  Inflation gave the spectrum of
fluctuations, but not the normalization.  The COBE-DMR
experiment measured the amplitude of temperature
fluctuations on an angular scale that 
was
acausal at last
scattering, and hence directly probed inflationary
fluctuations, and in particular the fluctuation strength.
More than twenty subsequent experiments have plugged the
causal gap, measuring fluctuations over angular scales less
than or of order the acoustic peak at
$\lambda=220\,\Omega^{1/2}$ that corresponds to the maximum
sound horizon in the early universe.

The shape of the fluctuation spectrum is now being probed.
The CMB fluctuations measure density fluctuations at $z \sim
1000$ on scales of hundreds of Mpc.  At large redshifts one
degree subtends a comoving scale of 100 Mpc.  A complementary
measure arises from galaxy redshift surveys.  These measure
variations in the luminous matter density out to $\sim 300$
Mpc at the present epoch.  One can combine, after choosing a
model, the CMB and 
large-scale structure
(LSS)
measures of the fluctuation spectrum.  
Here we will describe the current status of our
understanding of the shape of the primordial fluctuation
spectrum.

It is customary to use a two-parameter fit to the LSS power:
$\sigma_8$,
 the normalization at $8\,h^{-1}$ Mpc, and
$\Gamma$, measure of shape relative to CDM and 
nearly 
equal to $\Omega\,h$
for CDM.  Given the several data sets, each with a number of
independent data points, this may be an unnecessarily
restrictive approach.  Of course any data set is imperfect,
with possible systematic errors, and the data sets have
different selection biases.  However, 
provided the bias is
scale-independent, one can renormalize the data sets and
examine detailed shape constraints.  One has to decide
whether to compare a nonlinear power spectrum with the data
or whether to correct the data for nonlinearity and compare
the data with linear theory.  We will employ the latter
approach here.

\section{CMB:  Status of the Theory}

Inflation-generated 
curvature fluctuations provide the
paradigm for interpreting the cosmic microwave background
temperature anisotropies.  There are three components to
$\delta T/T$, schematically summarized as
$$\frac{\delta T}{T} = \vert \delta \phi + \delta
\rho_{\gamma} + \delta v \vert\,.$$
These are the gravitational potential, intrinsic and Doppler
contributions from the last scattering surface.  The
combined effect of the first two terms results in the
Sachs-Wolfe effect $\delta T/T=\frac{1}{3}\vert\delta\phi\vert$
which represents the only superhorizon contributions to
$\delta T/T$.  Inflationary initial conditions then require
that the Fourier component 
$\vert\delta T/T\vert_k \propto
\vert \cos (kv_s t_{ls})\vert$
 at the last scattering epoch
$t_{ls}$, where $k$ is the wavenumber and $v_s$ is the sound
speed, must  be constant as $k\rightarrow 0$.  Inflation of course 
specifies the phases of  density fluctuations that
decompose to sound waves of wavelength less than that of the
maximum sound horizon, $v_s t_{ls}\,$.  Longer wavelengths correspond
to power-law
growing modes at horizon crossing.  
The wave just
entering the horizon at last
scattering has a peak at wavenumber $n\pi/v_s t_{lss}$,
$n=1$, and a succession of waves crest at $n=2,\ 3, \dots$
before damping sets in as the photon mean free path increases
relative to the wavelength.  The first acoustic peak
projects to $\delta T/T$ on angular scales 
$\sim\,\Omega^{1/2}$ degree, and is a robust measure of the
curvature of the universe.  Doppler peaks are 90\deg out of
phase
 and of lower amplitude, so they fill in the troughs of the acoustic oscillations as 
measured by the radiation power spectrum.  Peak heights are determined in large part  by choice of
$\Omega_B$ and  $\Omega_\Lambda$. An
increase in $\Omega_B$  enhances the wave compression and reduces
the rarefaction phases.  An
increase in $\Omega_\Lambda$
enhances the ratio of radiation to matter in a flat
model, and thereby boosts the peak  potential decay and the low $\ell$
power via the early
integrated Sachs-Wolfe effect.  Of course increasing the
spectral index $n$ also raises the peak height.  Peak
heights are  lowered by reionization and secondary
scattering.  Not all of these degeneracies are removed by
examining the higher peaks.  For example, combination
of $\Omega_B$ and  $\Omega_\Lambda$ at specified $\Omega$ is 
nearly
degenerate
in peak height and peak location
since the angular size-redshift relation depends only on $\Omega.$

Lensing by nonlinear and
quasilinear foreground structure redistributes the peak
power towards very high $\ell$ in a way that breaks the
degeneracies in the CMB. One can search for this effect  with interferometer experiments \cite{met} at $\ell\simgt 1000$ or else by correlating the CMB fluctuations with large-scale power from either large redshift surveys such as
the SDSS \cite{wanss} or via weak lensing distortions of the CMB
\cite{selz} or LSS \cite{hut}.

\section{Reconstruction of the Primordial Power Spectrum
from the CMB}

The inflationary, approximately scale-free, power spectrum $P(k)=Ak^n$,
$n\approx 1$, is modified by the transition from radiation to matter domination,
since in matter-dominated epochs different growth occurs on subhorizon scales:
in the radiation era, there is no subhorizon growth of fluctuations.  This
modifies $P(k)$ to $P(k)\propto k^{-3}$ on larger scales.  The
transition occurs at the horizon during matter-radiation equality, namely
$12(\Omega h^2)^{-1}$ Mpc.  The Boltzmann equation can be  solved for temperature
fluctuations, mode by mode, and the solutions for $\delta T/T$ are  scaled
to agree
with the quoted errors for each CMB experiment.  For each specified cosmological
model, we then infer the power spectrum amplitude over scales that correspond to
the deprojection on the sky of the experimental window function
\cite{gw}.  We confirm
 that
standard CDM ($\Omega_{CDM} =1$, $h=0.5$)  fits the CMB data rather poorly, 
with best fit renormalization that corresponds to a high value of $\sigma_8$.
The $\Lambda$CDM model ($\Omega=1$, $\Omega_m=0.4$, $ h=0.6$) gives a
reasonably good fit  with an acceptable value of 
$\sigma_8.$  Much stronger constraints however
come when these fits are combined with LSS data.

\section{Reconstruction of $P(k)$ from Large-Scale Structure Data}

There are several large-scale structure data sets that one may use to
reconstruct $P(k)$.  Galaxy redshift surveys include the Las Campanas Survey of
25,000 galaxies
\cite{lcrs}, the PSC$z$ survey of 1,500 galaxies \cite{pscz}, and the SSRS2/CfA2 survey
of 7,000 galaxies \cite{ssrs2}.  There is also the APM cluster survey which probes to $\sim
300\,h^{-1}$ Mpc \cite{clusters}
and the real space inversion of the 2D APM galaxy survey
\cite{apm}.  The
local mass function of clusters 
\cite{vial}
has been used to measure
$\sigma_8\Omega^{0.6}$, 
and the high redshift cluster abundance  \cite{bah}
has been used to
break the degeneracy between $\sigma_8$ and $\Omega$.  
Peculiar velocities and
large scale bulk flows also yield $\sigma_8\Omega^{0.6}$  in a completely
bias-independent approach, although systematic uncertainties remain large
\cite{kold}.

The redshift space surveys can be corrected in a straightforward way, for
peculiar velocities on small scales and bulk flows on large scales, to derive
the real space $P(k)$ by assuming a cosmological model \cite{pead}.  
Correction of data in
the nonlinear regime
is best done by numerical simulation, but 
can be performed using
an empirical formulation 
calibrated to numerical simulations based on 
a smooth interpolation from spherical collapse by a factor of 2
in radius on cluster scales \cite{peadd}.  Renormalization of the various measures of $P(k)$
is effected by assuming that all measurements are subject to a scale-independent
bias, allowed to be independent for each probe of $P(k)$.

\section{Confrontation of $P(k)$ with CMB and LSS}

Model fitting to LSS alone results in the following conclusions.
  Of course
the standard COBE-normalized
CDM model fails completely.  Without a large scale-dependent bias factor on
10 -- 100 Mpc scales,  peculiar velocities and the galaxy cluster
abundance are greatly overpredicted.  Low density models circumvent these problems. The cluster abundance, evolution and baryon fraction are all
in satisfactory agreement with observations \cite{bah}.

However the combined CMB/LSS fits to $P(k)$ lead \cite{gw} to a surprising conclusion.
The surprise  is that almost all
 models, while occasionally faring better than sCDM, still provide
unacceptable fits to all of the data.  Consider for example $\Lambda$CDM,
currently favored by the SN Ia Hubble diagram.  The reduced $\chi^2$ is 
2.1 for
70 degrees of freedom.  Data set by data set, one still has a problem.  For
example the values of $\chi^2$(d.o.f.) 
are APM clusters:  25(8); LCRS 17(5); APM
44(9); IRAS 16(9).  The acceptable data sets are CfA for 2 d.o.f., cluster
abundances/peculiar velocities for 3 d.o.f. and CMB for 34 d.o.f.

\section{ Neutrinos and LSS}

 The only mildly
acceptable model (reduced $\chi^2 = 1.2$) is CHDM, hot and cold dark matter with
$\Omega=1$.  This model overpredicts current  cluster
abundances and  underpredicts the small number of high redshift, luminous
x-ray clusters (2 at $z=0.5$, 1 at $z=0.8$).
However the cluster evolution  constraint is disputed \cite{bla}, 
and
the local normalization
is not 
necessarily robust. 
The cluster baryon fraction provides an independent and 
powerful constraint that favors $\Omega_m\approx 0.3$.  
Of course this rests on the
reasonably
plausible assumption that clusters provide a fair sample of the baryon fraction of the universe. This need not necessarily be true if gas has had a complex history prior to cluster formation:
{\it e.g.} the gas  may have been preheated 
as is suggested by recent considerations of the entropy of intracluster 
gas \cite{bal}. This would reduce the baryon fraction, but one can equally well imagine scenarios for cluster formation in dense sheets or filaments where the baryon fraction was already enhanced.

Consider the model preferred by the combination of 
SNIa and cluster constraints, namely $\Lambda$CDM.
Figure 1
shows the $\Lambda$CDM power spectrum compared 
with observations of Large-Scale Structure and CMB anisotropy.
One can pose the following question: does adding a hot component improve the  marginally acceptable LSS fit?  We find that as an admixture of 
HDM is added to the dominant CDM, the combined fit to CMB and LSS deteriorates
%
(Figures 2,3).
The reason is that low
density
 CDM models have a 
$P(k)$ peak that is longward of the apparent peak in the APM data.
Adding HDM only 
exacerbates
the mismatch. 

Neutrino masses imprint  a distinct signature on $P(k).$ This will eventually be a measurable probe of the neutrino mass, from LSS as well as from CMB.  Indeed the LSS probe may potentially
be more powerful \cite{huet}.
 There is more dynamical
range available in probing $P(k)$ with LSS on the neutrino free streaming scale, where the primary signature should be present. Even present data is sensitive to a neutrino mass of around an eV: for example  we find that the fit changes significantly between 0.1 and 1 eV.  
  If $\Lambda$CDM is in fact the right model, our analysis yields an indirect
upper limit on the mass of the most massive neutrino species 
of $m_\nu \leq 2$eV.  While there are 
considerable systematic uncertainties in this approach, it is promising as a 
complement to the direct evidence for mass difference between neutrino 
species from SuperKamiokande \cite{fukuda} and the solar neutrino 
problem \cite{jbah}, and is already beginning to conflict with results from 
LSND \cite{ath} that require a large mass difference.  

One class of exotic models is the following.
Take a model that fits all constraints except for the shape. 
The best contender for such a model is $\Lambda$CDM.   
Inspection of the LSS constraints reveals that 
there is a deficiency of large-scale power near 100 Mpc. 
One can add an ad hoc feature  on this scale from considering 
inflationary models with
multiple scalar fields
(see Figure 4).
This could be generated for example by \cite{amendola}
incomplete coagulation of bubbles of new phase 
in a universe that already has been homogenized by a previous episode of
inflation. 
One can tune the bubble size distribution to be sharply peaked at any 
preferred
 scale. This results in nongaussian features and excess power where needed. 
The non-gaussianity provides a distinguishing characteristic.

 Other suggestions that fit both CMB and LSS data appeal to an inflationary  relic of excess power from broken scale invariance, arising from  double inflation  in a $\Lambda$CDM model,
which results in a gaussian feature  that is essentially  a step in $P(k)$ at the desired wavenumber \cite{les}. This 
 improves the fit in much the same way as adding a hot component to CDM 
 improves the empirical fit.
While such ad hoc fits may seem  unattractive, one could argue that 
other aspects of cosmological model building are equally ad hoc,
such as postulating a universe in which
$\Lambda$ is only  becoming dynamically important at the present epoch. 
Clearly one has to accommodate such arguments in order to  fit the data, 
if the data is indeed accepted at face value.
Moreover there are positive side effects that arise from the tuned void
approach.
The bubble-driven shells provide a source of overdensities on large scales. Rare shell interactions  could produce nongaussian  massive galaxies or  clusters at low or even high redshift: above a critical surface density threshold
gas cooling would help concentrate gas and aid collapse.
 If massive
galaxies were discovered at say $z>5$ or a massive galaxy cluster at $z>2$ this
would be another indication that the current library of cosmological models is
inadequate. New data sets such as SDSS and 2DF are 
urgently needed to verify whether the shape discrepancies in $P(k)$ will
persist.

\section{Summary}

If the data are accepted as mostly being free of systematics
and ad hoc additions to the primordial power spectrum are avoided,
 there is no acceptable model for
large-scale structure.  No one LSS data set can be blamed.  Perhaps it is best
to wait for improved data.  The Sloan and 2DF surveys are already acquiring
galaxy redshifts.  However another philosophy is to search for more exotic
models.  Consider for example the primordial isocurvature mode.  This has the
advantage of forming primordial black holes of stellar mass, since normalization
to large-scale structure and the present spectrum over 10 -- 50 Mpc requires 
a spectral index that generates nonlinear fluctuations at roughly the
epoch of the quark-hadron phase transition, when the horizon contained
approximately one solar mass \cite{sugi} 
(hence the primordial black holes may be the possibly observed  MACHOs). 
The goodness of fit of this model to
the combined CMB/LSS data is similar to that of the $\Lambda$CDM model. One cannot distinguish with current data between an exotic isocurvature model and
$\Lambda$CDM,  although neither model is satisfactory. To
improve on this, clearly something even more exotic is required.  It may
be that independent observations will force us in this direction. 

\section*{Acknowledgements}

We gratefully acknowledge support by NSF and NASA.

\section*{References}
%

\newpage
\pagestyle{empty}

\begin{figure}[htb]
\centerline{\psfig{file=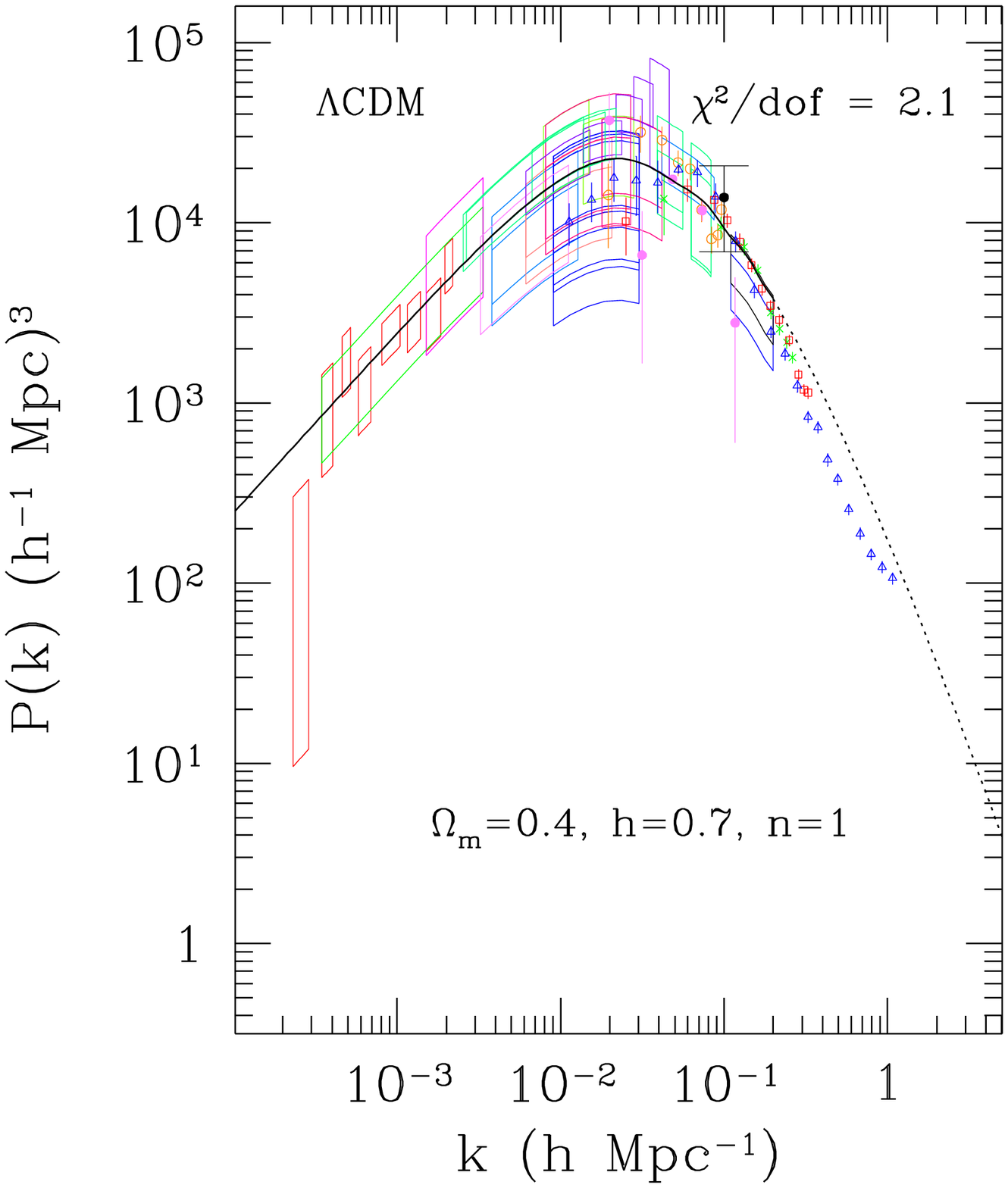,width=5in}}
\caption{Constraints from LSS and CMB on $\Lambda$CDM model} 
\end{figure}

\begin{figure}[htb]
\centerline{\psfig{file=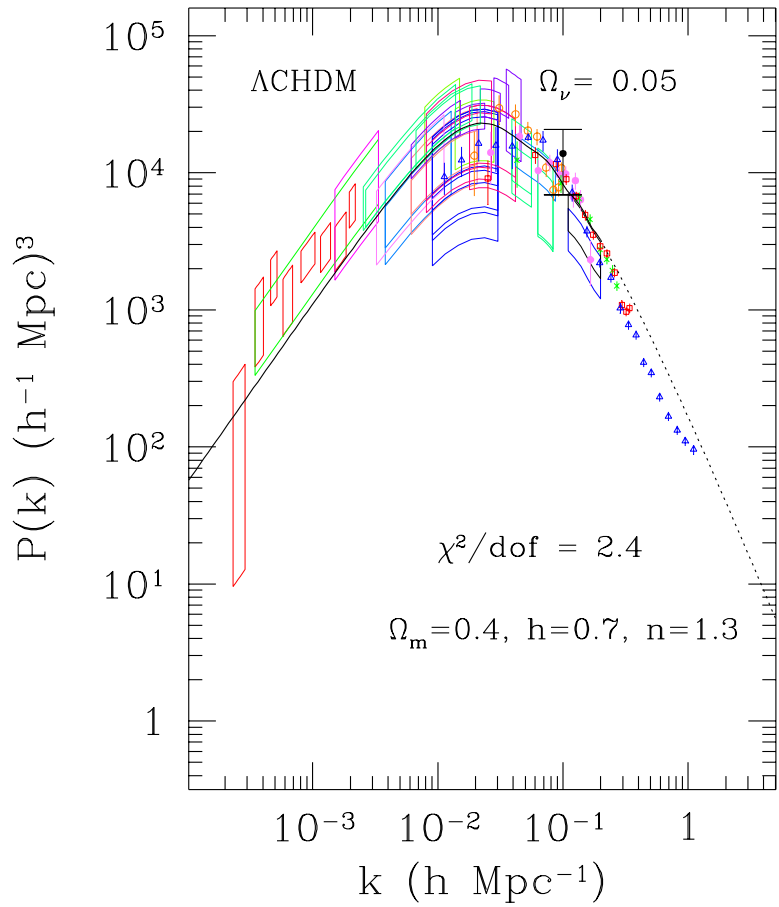,width=5in}}
\caption{$\Lambda$CDM model with $\Omega_\nu=0.05$.  
A blue tilt of the 
primordial power spectrum ($n=1.3$) 
is necessary to counteract the damping 
of small-scale perturbations by free-streaming of the 
massive neutrinos, which makes the peak of the model fall 
even farther below that of the data unless $n>1$.
Even with this best-fit value of $n$, the 
fit to the data is worse than with no HDM, because 
CMB observations disfavor such a high value of $n$.  
}
\end{figure}

\begin{figure}[htb]
\centerline{\psfig{file=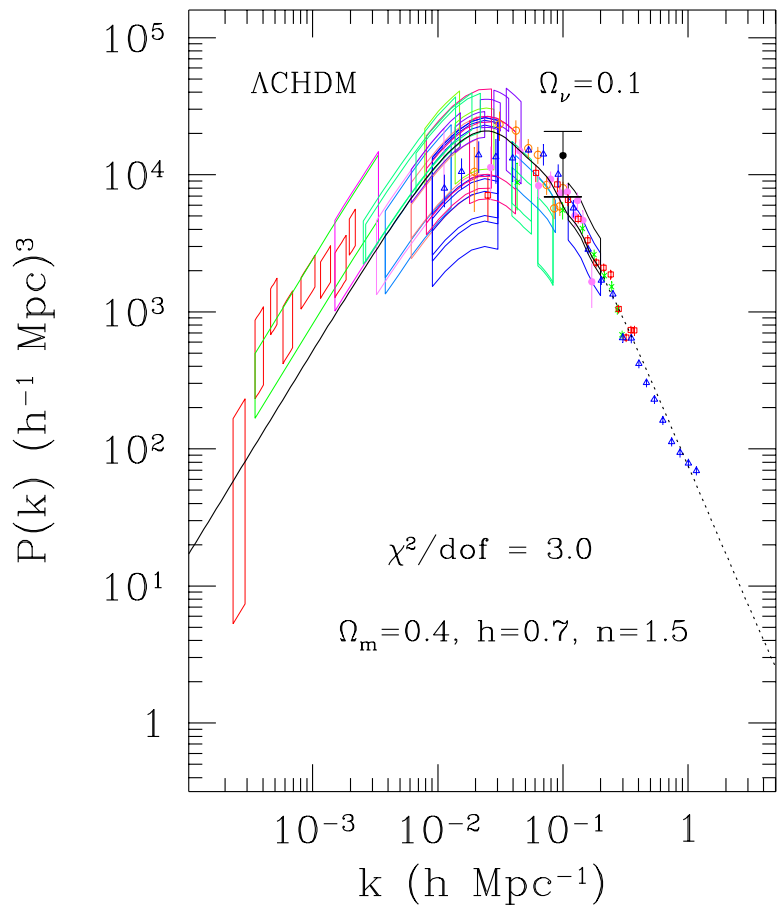,width=5in}}
\caption{$\Lambda$CDM model with $\Omega_\nu=0.10$.  The best-fit 
value of $n$ is now 1.5.  The fit 
to the data worsens as more HDM is added.}  
\end{figure}

\begin{figure}[htb]
\centerline{\psfig{file=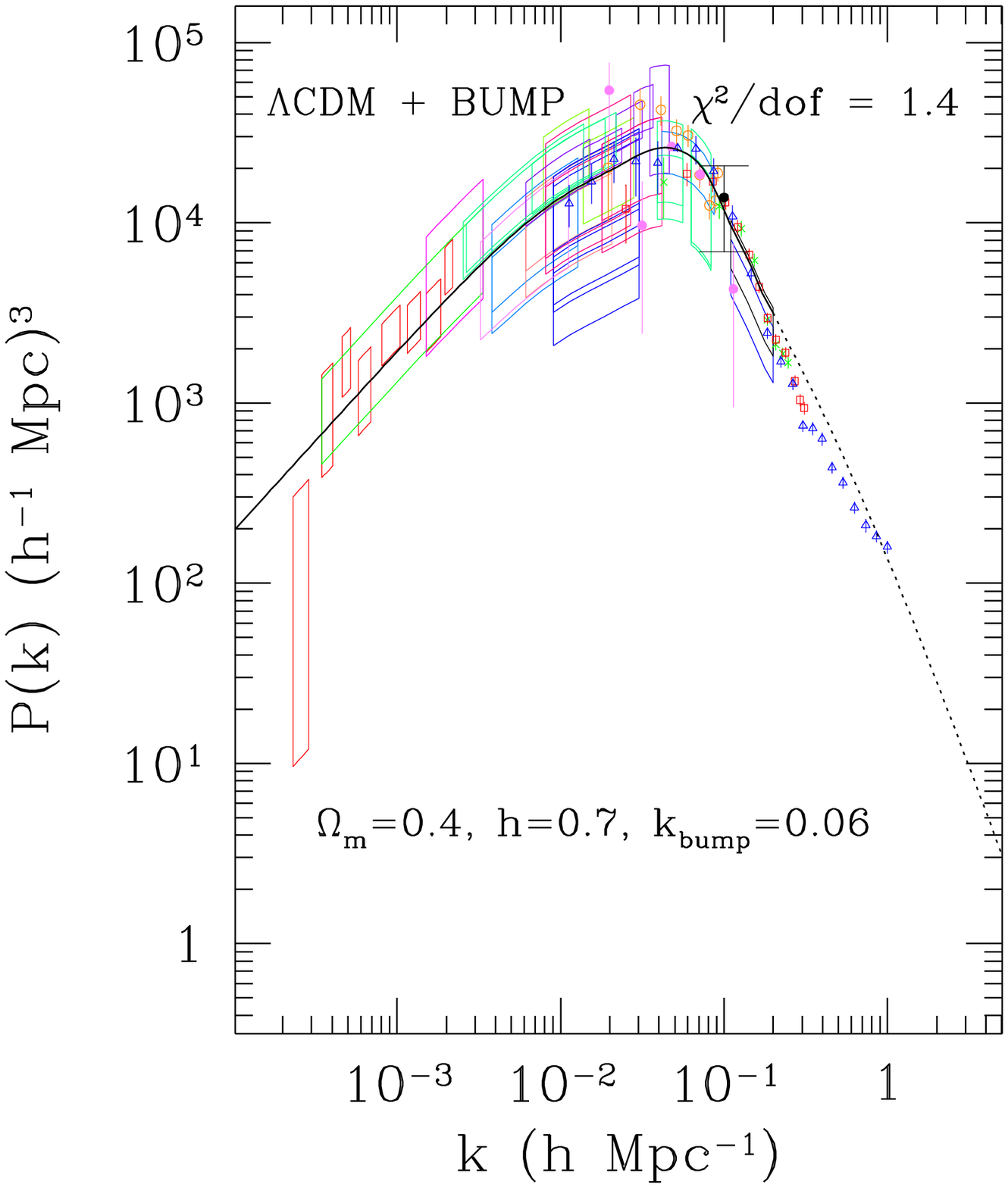,width=5.0in}}
\caption{Constraints from LSS and CMB on $\Lambda$CDM model with 
a broad enhancement centered at $k=0.06h^{-1}$Mpc 
added to the primordial power spectrum.}
\end{figure}


\begin{thebibliography}{99}
%
\bibitem{met}  B. Metcalf and J. Silk, ApJ {} {492L}{1} {(1998)}
\bibitem{wanss}  Y. Wang, D. N. Spergel and M. A. Strauss,
 preprint astro-ph/9802231, ApJ, in press
(1999)
\bibitem{selz}  M. Zaldarriaga and U. Seljak,  preprint astro-ph/9810257
(1998)
\bibitem{hut} W. Hu and M. Tegmark, preprint astro-ph/9811168 (1998)
\bibitem{amendola} L. Amendola, C. Baccigalupi, R. Konoplich, F. 
Occhionero and S. Rubin, Phys. Rev. D {} {54} {7199} {(1996)}
%
\bibitem{gw} E. Gawiser and J.  Silk, Science {} {280} {1405} {(1998)}
\bibitem{lcrs} H. Lin et al., ApJ {} {471} {617} {(1996)}
\bibitem{pscz} W. Saunders et al., Proc. of {\it XVIIth Rencontres de 
Moriond:  Extragalactic Astronomy in the Infrared}, ed. G. A. Mamon, T. X. 
Thuan, J. T. T. Van, (1997), p.431
\bibitem{ssrs2} L. N. Da Costa et al., ApJ {} {437} {L1} {(1994)}
\bibitem{clusters} H. Tadros, G. Efstathiou, G. Dalton, MNRAS {296} {995} {(1998)}
\bibitem{apm} E. Gazta\~naga and C. M. Baugh, MNRAS {} {294} {229} {(1998)}
\bibitem{vial} P. T. P. Viana and A. R. Liddle, MNRAS {} {281} {323} {(1996)}
\bibitem{bah} N. A. Bahcall, X. H.  Fan and  R. Y. Cen,   ApJ {} {485L} {53} 
{(1997)}
\bibitem{kold} T. Kolatt and A. Dekel, ApJ {} {479} {592} {(1997)}
\bibitem{pead} J. A. Peacock and S. J. Dodds, MNRAS 
{267} {1020} (1994).
\bibitem{peadd} J. A. Peacock and S. J. Dodds, MNRAS 
{280} {L19} (1996).
\bibitem{bla} A. Blanchard, J. Bartlett and R. Sadat  preprint astro-ph/9809182
(1998)
\bibitem{bal} M. Balogh, A. Babul and D. R. Patton, preprint astro-ph/9809159
(1998)
\bibitem{huet} W. Hu, D. Eisenstein, and M. Tegmark, Phys. Rev. Lett., {}
{80} {5255} {(1998)}
\bibitem{fukuda} Y. Fukuda et al., preprint hep-ex/9805021, Phys. Rev. Lett. 
in press (1998)
\bibitem{jbah} J. N. Bahcall, P. I. Krastev, and A. Yu. Smirnov, preprint hep-ph/9807216, (1998)
\bibitem{ath} C. Athanassopoulos et al., Phys. Rev. Lett. {} {77} {3082} {(1996)}
\bibitem{les} J. Lesgourgues, D. Polarski and A. A. Starobinsky, MNRAS {297} {769} (1998)
\bibitem{sugi} N. Sugiyama and J. Silk, preprint (1998)


\end{thebibliography}
\end{document}